\documentclass[aps,prd,twocolumn,notitlepage,preprintnumbers,nofootinbib,longbibliography]{revtex4-1}
\usepackage[utf8]{inputenc}
\usepackage{amsmath}
\usepackage{amssymb}
\usepackage{color}
\usepackage{braket}
\usepackage{slashed}
\usepackage{xcolor}
\usepackage{graphicx}
\usepackage{mathrsfs}
\usepackage{subfig}
\usepackage{tikz}
\usepackage[pdfpagelabels=false,colorlinks=true,urlcolor=teal,citecolor=teal]{hyperref}

\newcommand{\rd}{\mathrm{d}}

\newcommand{\nn}{\nonumber}

\newcommand{\e}{\epsilon}
\newcommand{\rT}{\mathcal{T}}

\newcommand{\cO}{\mathcal{O}}

\definecolor{darkred}{rgb}{0.7,0.0,0.0}
\definecolor{darkblue}{rgb}{0.0,0.0,0.9}
\definecolor{darkgreen}{rgb}{0.0,0.5,0.0}
\definecolor{brown}{rgb}{0.0,0.0,0.0}
\newcommand{\red}{\color{darkred}}
\newcommand{\blue}{\color{darkblue}}
\newcommand{\green}{\color{darkgreen}}

\newcommand{\cyan}{\green}

\def\ib{{\; \bar\imath}}

\newcommand{\intlim}[3]{\int_{#1}^{#2}\! \rd #3 \,}

\begin{document}
\preprint{MPP-2025-158}
\title{Energy Correlators in Semi-Inclusive Electron-Positron Annihilation}

\author{Yu Jiao Zhu}
\email{yzhu@mpp.mpg.de}
\affiliation{Max-Planck-Institut f\"{u}r Physik, Werner-Heisenberg-Institut, Boltzmannstr. 8, 85748 Garching, Germany}

\begin{abstract}
We investigate energy correlators  in semi-inclusive electron-positron annihilation as precision probes of parton-to-hadron transition. 
Using soft-collinear effective theory, we analyze the correlation patterns between the examined hadron and the  accompanying QCD radiation
 in both large-angle and small-angle limits.
 We derive complementary factorization formulas in the back-to-back and collinear limits, 
 involving TMD fragmentation functions and fragmentation energy correlators, respectively. 
 We compute the NNLO twist-two matching coefficients of the latter and obtain renormalization-group-improved predictions at N\(^3\)LL accuracy in the back-to-back region and NNLL accuracy in the collinear region.
Our results demonstrate that semi-inclusive energy correlators provide direct, theoretically controlled access to QCD dynamics underlying hadronization, 
opening new avenues for precision studies at future lepton colliders as well as through reanalyses of archival LEP data.
\end{abstract}

\maketitle
In high-energy collisions, quarks and gluons produced at short distances cannot be directly observed due to color confinement. Instead, they undergo a nonperturbative transition process known as hadronization, in which colored partons transform into color-neutral hadrons. This transition, governed by the long-distance dynamics of QCD, bridges the calculable partonic cross sections and the experimentally accessible hadronic final states.
Among various processes, the single-inclusive hadron production in electron-positron annihilation (SIA), $\ell +  \ell' \to  h  + X$,
offers the cleanest probe of parton-to-hadron  transition.
The nonperturbative aspects of this transition are encoded in fragmentation functions, which describe the probability for a parton to produce a specific hadron carrying a given fraction of its momentum, and enter the cross section through a convolution with perturbatively calculable coefficient functions~\cite{Collins:2011zzd,Collins:1981ta,Bodwin:1984hc,Collins:1985ue,Collins:1988ig,Collins:1989gx,Nayak:2005rt}, which are known up to next-to-next-to-next-to-leading order (N\textsuperscript{3}LO) in QCD~\cite{Rijken:1996ns,Mitov:2006wy,He:2025hin}, including threshold resummation effects~\cite{Xu:2024rbt}.

In addition to the conventional SIA cross section, the energy correlators (ECs) have emerged as a complementary probe of hadronization, mapping partonic correlations in the microscopic energy flux onto the observed macroscopic correlations among hadrons, and introducing new nonperturbative correlation functions~\cite{Liu:2022wop,Liu:2024kqt,Lee:2025okn,Chang:2025kgq,Kang:2025zto}.

The energy-energy  correlators were first introduced in Refs.~\cite{Basham:1977iq,Basham:1978bw,Basham:1978zq,Basham:1979gh} as  tests of strong-interaction dynamics in the early development of QCD. They can be expressed as correlation functions of energy-flow operators
\begin{align}
\label{eq:eec-def-0}
\langle \mathcal{E}(\hat n_1) \mathcal{E}(\hat n_2) \rangle
= \int \rd^4 x \, e^{i q\cdot x}
\langle 0| J(x) \mathcal{E}(\hat n_1) \mathcal{E}(\hat n_2) J(0) |0 \rangle\,,
\end{align}
where $J$ denotes the source current, e.g., the electromagnetic current $J^\mu = \bar{\psi} \gamma^\mu \psi$.
ECs have since been generalized to diverse collision systems, enabling the study of transverse-momentum-dependent dynamics via back-to-back jets~\cite{Collins:1981zc,deFlorian:2004mp,Moult:2018jzp,Duhr:2022yyp,Aglietti:2024xwv,Korchemsky:2019nzm,Chen:2023wah,Moult:2019vou,Kardos:2018kqj,Tulipant:2017ybb,vonKuk:2024uxe,Aglietti:2024zhg,Kang:2024dja,Li:2020bub,Gao:2019ojf,Li:2021txc,Gao:2023ivm} and jet substructure via the collinear limit~\cite{Lee:2024icn,Lee:2024tzc,Liu:2024lxy,Dokshitzer:1999sh,Lee:2024esz,Chen:2024nyc,Konishi:1978ax,Konishi:1979cb,Konishi:1978yx,Dixon:2019uzg,Chen:2019bpb,Chen:2023zzh,Craft:2022kdo,Chen:2021gdk,Chen:2020adz,Chen:2025rjc,Chen:2023zlx,Jaarsma:2023ell,Budhraja:2024tev}. They have been further applied to probe quarkonium production~\cite{Chen:2024nfl}, the nucleon  spin structures~\cite{Kang:2023big,Kang:2023gvg} and small-$x$ dynamics~\cite{Mantysaari:2025mht,Liu:2023aqb,Bhattacharya:2025bqa,Kang:2025vjk,Kang:2023oqj}. 
Moreover, 
the rigorous operator definition of the energy correlators offers a clear path toward a first-principle  understanding of underlying nonperturbative dynamics.
For instance, universality in nonperturbative corrections has been investigated both theoretically and through global fits~\cite{Belitsky:2001ij,Korchemsky:1994is,Korchemsky:1997sy,Korchemsky:1999kt,Schindler:2023cww,Lee:2006nr,Lee:2006fn,Abbate:2010xh,Herrmann:2025fqy,Cuerpo:2025zde}, and recent measurements across $e^+e^-$ and hadronic collisions~\cite{ALICE:2024dfl,STAR:2025jut,Bossi:2024qeu,Bossi:2025xsi} have opened new opportunities to test these predictions. See Ref.~\cite{Moult:2025nhu} for a comprehensive review.

Motivated by the need to access hadronization dynamics, 
we introduce the semi-inclusive energy  correlator  as a variant of the inclusive EEC in Eq.~(\ref{eq:eec-def-0})
\begin{align}
\langle \mathcal{E}(\hat n)  \rangle_h
= \int \rd^4 x \, e^{i q\cdot x}
\langle 0| J(x) \mathcal{E}(\hat n) a_h^\dagger a_h  J(0) |0 \rangle\,,
\end{align}
where  $h$ denotes  an  identified hadron created by  $a_h^\dagger$, 
and  $\hat n$  is the   generic radial unit vector where energy weight is applied.
We study its dependence 
on   the relative polar angle $\chi$  between $\hat n$ and $\hat n_h$
both in the small- and wide-angle limits, 
where effective field theory is applied to reveal the corresponding factorization structures,
and obtain
precision predictions  
via
  renormalization group (RG) evolution.
\section{ECs in  single inclusive hadron production}
In this paper, we consider polar angle $\chi$ distribution for single inclusive  hadron production  in electron-positron annihilation
\begin{align}
 \ell(p_\ell) +  \ell'(p_{\ell'}) \to \gamma^\ast(q)\to  \pi(P_\pi)  + X(P_X)
\,,\end{align}
where the identified hadron is taken as a pion.
We  work in the overall center-of-mass frame, in which the virtual photon is at rest
$
q=\frac{Q}{2}(\bar n +n)=Q (1\,,0\,,0\,,0)\,$.
We will  parametrize the events by angular variable $\chi$,
defined  to be the polar angle between the detected particles (jets) with respect to  pion velocity.

We are interested in energy weighted angular $\chi$ correlations 
\begin{align}
E_\pi\frac{\rd \sigma^\text{EC}}{\rd^3\vec P_\pi}=\frac{2\alpha_e^2}{Q^6} L_{\mu\nu}W_\text{EC}^{\mu\nu}\,.
\end{align}
 In this case, the conventional  hadronic tensor~\cite{Collins:2011zzd}  is replaced by a correlator in which the energy-flow operator is inserted between the electromagnetic currents
\begin{align}
W_\text{EC}^{\mu\nu}(q,P_\pi)=
&
\frac{1}{4\pi}
\sum_X
\int
\rd^4 z
\,
e^{i q\cdot z}
\nn\\
\times&\langle 0 |
j^\mu(z/2)
\hat{\mathcal{E}}(\chi)
| P_\pi\,,X\rangle 
\langle P_\pi\,,X| j^\nu(-z/2) |0\rangle\,,
\end{align}
where the cumulant EC operator quantifies the energy deposited in the detector for radiation within a radial angle bounded above by $\chi$ 
\begin{align}
\label{eq:eec-def}
\hat{\mathcal{E}}(\chi)|X\rangle = \sum_{h \in X}\frac{E_h}{E_\pi}\Theta(\chi-\chi_{h-\pi})|X\rangle\,.
\end{align}
The  spin-averaged leptonic tensor $L^{\mu\nu}$  for the  decay of a virtual photon is given by  
\begin{align}
L^{\mu\nu}=  p_\ell^\mu p_{\ell'}^\nu+p_\ell^\nu p_{\ell'}^\mu -\frac{Q^2}{2} g^{\mu\nu}\,.
\end{align}
Using SO(3) global rotation symmetry, we  integrate over all polar directions, this leaves behind a single  kinematic invariant
\begin{align}
x=\frac{2 P_\pi\cdot q}{ Q^2}\,,
\end{align}
which is  related to the Deep-Inelastic Scattering (DIS) Bjorken $x_B$ by inversion $x=1/x_B$,
$x$ is interpreted as the center-of-mass energy of the detected pion relative to its maximum value $Q/2$.
The  energy weighted cross section  receives combined contribution from longitudinal and transverse   polarization modes 
\begin{align}
\label{eq:structure-fun}
\rd  \sigma^{\text{EC}}(x,Q, \chi) = \rd  \sigma_L^{\text{EC}}(x,Q, \chi)  + \rd  \sigma_T^{\text{EC}}(x,Q, \chi)\,.
\end{align}
We will also investigate Bjorken $x$ weighted ECs, defined as follows
\begin{align}
\label{eq:mellin-EEC}
 \sigma^\text{EC}_\pi(N,Q,\chi) 
= &\sum_h \int  x^{N-1}\rd \sigma(x,Q, P_h) \frac{E_h}{E_\pi}\Theta(\chi-\chi_{h-\pi})
\nn\\
= &\int  x^{N-1} \rd \sigma_\pi^{\text{EC}}(x,Q, \chi)\,.
\end{align}
\section{Back-to-Back Limit and the TMD Fragmentation  Functions}
\begin{figure}
\centering
\includegraphics[width=0.25 \textwidth]{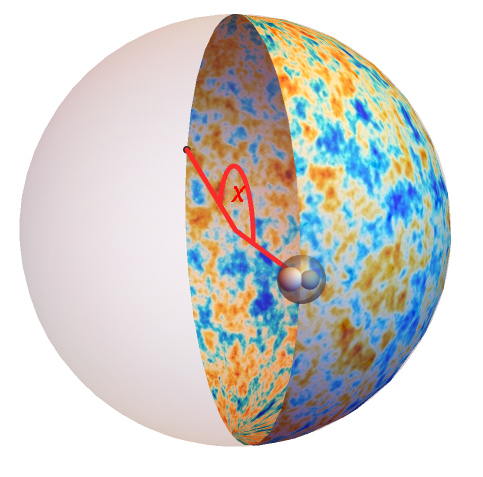}
\caption{ECs back-to-back limit.}
\label{fig:EEC_cmb_pi}
\end{figure}
In this section, we are concerned with ECs in the back-to-back limit $\chi\to\pi$, see Fig.~\ref{fig:EEC_cmb_pi}. 
The angular  $\chi$ correlation  is  closely related to the conventional $q_T$ spectrum,
the latter measures  the transverse momentum of the detected  hadrons in the photon-pion frame, 
or the photon transverse momentum in hadron-pion frame.
The explicit relation is, at leading power in the back-to-back limit
\begin{align}
\sin\chi=\frac{q_T}{Q/2}\,.
\end{align}
This kinematic region is  understood  as the crossed counterpart  of current-fragmentation region (CFR) in semi-inclusive deep-inelastic scattering (SIDIS),
and the appropriate theoretical framework for describing it is transverse momentum dependent (TMD)  factorization.
To derive the factorization formulas, it is useful to identify a hadron inside the leading jet in the first place.
To this end, we consider  differential cross section with two identified hadrons
\begin{align}
\rd \sigma_{\ell+ \ell'\to h_A+h_B+X}=
&
\frac{1}{2Q^2}\frac{(4 \pi \alpha_e)^2}{Q^4}N_c
\prod_{i=A,B}\frac{\rd ^3 P_i}{(2\pi)^32E_i}
\nn\\
\times
&
 L_{\mu \nu}(P_\ell,P_{\ell'}) W^{\mu \nu}(q, P_A, P_B)\,.
\end{align}
The hadronic tensor with two resolved hadrons is 
\begin{align}
&W^{\mu\nu}(q,P_A,P_B)=\prod_X \frac{\rd^3 P_X}{(2\pi)^3 2 E_X} 
(2\pi)^4\delta^{(4)}(P_A+P_B
\nn\\
&+ P_X-q) \times
\langle 0 |
j^\mu
| P_A\,,P_B\,, X\rangle 
\langle P_A\,,P_B\,, X| j^\nu |0\rangle\,.
\end{align}
Kinematic constraints in  the back-to-back require that all final-state radiation in $X$ be collimated, with momenta  collinear to  either of the observed hadrons, and the hadronic amplitude is factorized through~\cite{Feige:2014wja}
\begin{align}
\langle h_{\blue{A}} h_{\green{B}};    {X_{{\blue{A}}\green{B}}}; &X_{\red{ \text {s}}}|J^\mu|0\rangle\simeq 
\,\mathcal{C}(Q^2,\mu)
 \langle X_{\red{ \text {s}}}| \rT \left[
  Y^\dagger_{\green{\bar n}}Y_{{\blue{n}}}
  \right]
  |0\rangle
  \nn\\
  \times &
\langle h_{\green{B}}; {X_{\green{B}}}|\bar\chi_{\green{{\bar n}}}|0\rangle
\gamma^\mu
\langle h_{\blue{A}} ; {X_{\blue{A}}}| \chi_{\blue{n}}| 0\rangle\,,
\end{align}
where we  absorb  vacuum expectation values of zero-bin Wilson lines into the normalization of the collinear fields,
this is  equivalent  to the zero-bin subtracted soft-collinear eﬀective theory (SCET) Lagrangian 
~\cite{Feige:2014wja,Lee:2006nr}.
The hard amplitude $\mathcal{C}(Q^2,\mu)$ is  insensitive to the  infrared (IR) dynamics of   soft and collinear degrees of freedom,
and thus can be computed at partonic level. 
It is advantageous to consider $\gamma^\ast \to q \bar q$, in the absence of real soft radiation
\begin{align}
\label{eq:scet-qqb}
\langle {\green{q}} {\blue{\bar q}}|J^\mu| 0\rangle = 
\,\mathcal{C}(Q^2,\mu)
\langle {\green{q}}|\bar\chi_{\green{{\bar n}}}|0\rangle
\gamma^\mu
\langle {\blue{\bar q}}| \chi_{\blue{n}}| 0\rangle\,.
\end{align}
In this case, the leading-power collinear expansion is exact ($=$ instead of $\simeq$), 
as each  collinear trajectory contains  just one particle--taking the collinear limit simply does nothing.
By construction, the IR divergences of the partonic amplitude $\langle {\green{q}} {\blue{\bar q}}|J^\mu| 0\rangle$
is reproduced by the UV divergences of the hard scattering operators in SCET~\cite{Bauer:2000ew,Bauer:2000yr,Bauer:2001ct,Bauer:2001yt}.
 Although  a thorough treatment would also require 
  Glauber operators   to reproduce the  $i\pi$ terms in the low-energy part of the amplitude $\gamma^\ast \to q \bar q$,
the Cheshire nature of the Glauber exchange ensures that 
   the hard matching coefficients remain unchanged without including the Glauber interactions~\cite{Rothstein:2016bsq}.
Indeed,  the right-hand side of Eq.~(\ref{eq:scet-qqb}) is
\begin{align}
\label{eq:pole-transmu}
  \langle {\cyan  q} {\blue \bar q}| \cO^{\vec \lambda}_\text{scet} | 0\rangle
 =  \langle {\cyan  q} {\blue \bar q}| \cO^{\vec \lambda}_\text{scet} |0\rangle_\text{tree}\left(1+\delta_{Z_{\vec \lambda}} (\epsilon_\text{IR})\right)\,,
\end{align}
where  the helicity basis is $\cO^{\vec \lambda}_\text{scet}= \bar\chi^\pm_{\green{{\bar n}}} \gamma^\mu \chi^\pm_{\blue{n}}\,,\,\,\, \lambda=\pm$\,.
The result is  proportional to tree-level 't Hooft-Veltman~\cite{tHooft:1972tcz} (tHV) bases because pure virtual-loop corrections in SCET
are scaleless and vanish identically. 
As a result, the IR poles of a  helicity amplitude are reproduced
 by UV renormalization counterterm diagrams provided by $\delta  {Z_{\vec \lambda}}= {Z_{\vec \lambda}}-1$~\cite{Moult:2015aoa}.
 
Once the hard and soft-collinear degrees of freedom are disentangled, 
the next step is to enumerate  all allowed hard events,
and within each hard label, zoom in to the soft-collinear subprocesses and count over  particles  in a proper way:
the classes of momenta entering into the soft-collinear phase space 
differ only by their typical size of energy or rapidity, as a result, the phase space should be integrated over with a rapidity cutoﬀ $\nu$,
here we employ exponential cutoﬀ  scheme~\cite{Li:2016axz}.
Additionally,  a zero-bin subtraction is performed to remove double-counting between collinear and soft sectors,
yielding genuine TMD fragmentation functions~\cite{Manohar:2006nz}. Consequently, we have factorization formulas~\cite{Collins:2011zzd}
\begin{widetext}
\begin{align}
&\frac{\rd \sigma_{\ell+ \ell'\to h_A+h_B+X}}{\rd z_A \rd z_B \rd \cos\theta\, \rd \vec q_\perp}
\simeq 
\frac{N_c}{2} \frac{4 \pi \alpha_e^2}{Q^4}
L_{\mu \nu}
\sum_f
H_f(Q^2\,,\mu)
\int 
 \frac{\rd\vec b_\perp}{(2\pi)^2}
 e^{-i\vec b_\perp\cdot \vec q_\perp}
  \mathcal{S}(b_\perp,\mu,\nu)
   \mathrm{Tr}\left [
   \mathscr{F}_{ h_{\blue{A}}/\bar {f}}
\gamma^\mu
\mathscr{F}_{ h_{\green{B}}/f}
\gamma^\nu
 \right  ]
   \nn\\
   =&
    \frac{4 \pi \alpha_e^2}{Q^2}
   N_c \frac{1+\cos^2\theta}{8}
    \sum_f
H_f(Q^2\,,\mu)
\int 
 \frac{\rd\vec b_\perp}{(2\pi)^2}
  e^{-i\vec b_\perp\cdot \vec q_\perp}
  \mathcal{S}(b_\perp,\mu,\nu)
    \mathscr{F}_{ h_{\blue{A}}/\bar {f}}\left(z_A,\frac{b_\perp}{z_A},
E_{n},\mu,\nu \right)
\mathscr{F}_{ h_{\green{B}}/f}\left(z_B,\frac{b_\perp}{z_B},
E_{\bar n},\mu,\nu\right)\,,
\end{align}
\end{widetext}
where  
$H_f(Q^2,\mu)$ 
is  the squared timelike continuation of the SIDIS hard matching coefficient~\cite{Becher:2006mr,Baikov:2009bg,Gehrmann:2010tu},
 $z_i$ are  momentum  fractions, and $E_n$ ($E_{\bar{n}}$) are the energies of  fragmented  partons.
$\vec q_\perp$ is the transverse momentum of the virtual photon in hadron-hadron frame,
or is a parametrization of the net transverse momentum of the detected  dijets $\vec q_\perp=-\sum_{i=A,B}\vec P_i^\perp/z_i$ as a result of soft recoil~\cite{Collins:2011zzd}.
 $\theta$ is the angle between the back-to-back jets and the lepton beam, which we choose to integrate over.
 Consequently
  \begin{widetext}
\begin{align}
\label{eq:fac-dihadron}
\frac{\rd \sigma_{\ell+ \ell'\to h_A+h_B+X}}{\rd z_A \rd z_B  \rd \vec q_\perp} \simeq 
\frac{4 \pi \alpha_e^2}{3Q^2}N_c
 \sum_q
H_q(Q^2\,,\mu)
\int 
 \frac{\rd\vec b_\perp}{(2\pi)^2}
  e^{-i\vec b_\perp\cdot \vec q_\perp}
 D^{\bar q}_1\left(z_A\,,\frac{b_\perp}{z_A}\,,\xi^{ n}, \mu\right)
 D^q_1\left(z_B\,,\frac{b_\perp}{z_B}\,,\xi^{\bar n}, \mu\right)\,,
\end{align}
\end{widetext}
where the physical TMDs are obtained by  dressing the genuine collinear functions   with the cloud of soft gluons
\begin{align}
D^q_1\left(z\,,\frac{b_\perp}{z}\,,\xi, \mu\right)=& \,\mathcal{F}_{h/q}\left(z\,,\frac{b_\perp}{z}\,,E, \mu,\nu\right) \sqrt{\mathcal{S}(b_\perp,\mu,\nu)}\,.
\end{align}
As a result, the rapidity divergences $\ln(4 E_i^2/\nu^2)$ cancel between collinear and soft sectors,
 leaving  behind physical rapidity logarithms $\ln(\xi^i/\mu_b^2)$ with $\mu_b=b_0/b_T, b_0=2e^{-\gamma_E}$.
$\xi^i$ are the Collins-Soper (CS) scales~\cite{Collins:1981uk,Collins:1987pm,Collins:2011zzd}.
 These rapidity scales are subject to the constraint  $\xi^n\xi^{\bar{n}}=Q^4$.
For dihadron production, a canonical choice is
$
\xi^n = \xi^{\bar n} = Q^2\,.
$
In this work, we are concerned with energy correlations between a single identified hadron and the  jet in the opposite direction.
By summing and integrating over the phase space of  one of the dihadrons, 
we recover a factorization  formula  given by the convolution of TMDs with EEC TMD jet function
 \begin{widetext}
\begin{align}
\label{eq:eec-hadron}
z\frac{\rd \sigma^\text{EC}_{\ell+ \ell'\to h+X}}{\rd z  \,\rd \vec q_\perp} \simeq 
\frac{4 \pi \alpha_e^2}{3Q^2}N_c
 \sum_q
H_q(Q^2\,,\mu)
\int 
 \frac{\rd\vec b_\perp}{(2\pi)^2}
  e^{-i\vec b_\perp\cdot \vec q_\perp}
 J_{\bar q}\left(b_\perp\,,\xi^{ n}, \mu\right)
  D^q_1\left(z\,,\frac{b_\perp}{z}\,,\xi^{\bar n}, \mu\right)\,,
\end{align}
\end{widetext}
where we perform twist-2 matchings of the TMD fragmentation function 
\begin{align}
D_1^q\left(z,\frac{b_\perp}{z}\,,\xi\,,\mu\right)
=&\sum_i  d_{h/i} \otimes \mathcal{C}_{iq}+ \mathcal{O}(b_T^2\Lambda^2_{\text{QCD}}) \,,
\end{align}
and  use FF sum rules to obtain the EEC TMD jet function~\cite{Moult:2018jzp}
\begin{align}
J_q(b_\perp\,,\xi,\mu)
\equiv&
\sum_i\int_{0}^{1}\rd \hat z \,\hat z\, {\mathcal{C}}_{iq}\left( \hat z,\frac{b_\perp}{\hat z}\,,\xi\,,\mu\right)\,.
\end{align}
The factor $z$ in the front of Eq.~(\ref{eq:eec-hadron})  arises due to $E_h/E_\pi =1/x \, E_h/(Q/2)$, and  $x=z$ in the  back-to-back limit.
By summing over nonordered tuples $\sum_{\{A,B\}}=1/2 \sum_B  \sum_A$, the factorization formula for  energy weighted   back-to-back jets reads 
\begin{figure}
\centering
\includegraphics[width=0.32 \textwidth]{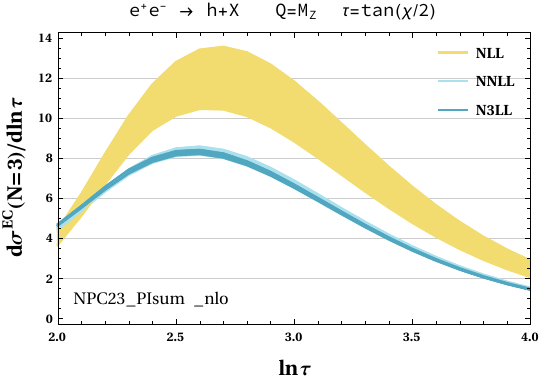}
\caption{Semi-inclusive ECs without $b_\ast$  prescription~\cite{Collins:2014jpa} in the back-to-back limit,
with   \texttt{LHAPDF6}~\cite{Buckley:2014ana} FF  dataset  \texttt{NPC23-PIsum-nlo}~\cite{Gao:2024nkz,Gao:2024dbv}.}
\label{fig:EEC-btob}
\end{figure}
\begin{widetext}
 \begin{align}
\label{eq:eec-jet}
\frac{\rd \sigma^\text{EEC}_{\ell+ \ell'\to X}}{\rd \vec q_\perp} 
\simeq&\sum_{\{A,B\}} \int  \prod_{i=A,B}  z_i  \rd z_i  \frac{\rd \sigma_{\ell+ \ell'\to h_A+h_B+X}}{\rd z_A \rd z_B  \rd \vec q_\perp} 
=
\frac{1}{2}
\sum_h \int z^2 \rd z 
\frac{\rd \sigma^\text{EC}_{\ell+ \ell'\to h+X}}{\rd z  \,\rd \vec q_\perp} 
\nn\\
= & 
\,\frac{4 \pi \alpha_e^2}{3Q^2}N_c
\frac{1}{2}
 \sum_q
H_q(Q^2\,,\mu)
\int 
 \frac{\rd\vec b_\perp}{(2\pi)^2}
  e^{-i\vec b_\perp\cdot \vec q_\perp}
 J_{\bar q}\left(b_\perp\,,\xi^{ n}, \mu\right)
  J_{ q}\left(b_\perp\,,\xi^{ \bar n}, \mu\right)\,.
\end{align}
\end{widetext}
The above analysis is  consistent  with Ref.~\cite{Moult:2018jzp} and reproduces the fixed-order result of Ref.~\cite{Dixon:2018qgp} in dijet limit.
Nonperturbative corrections to back-to-back EEC  have recently been analyzed  in Refs.~\cite{Kang:2024dja,Cuerpo:2025zde}.

The $\mu$-evolution  of the hard matrix and the TMDs is linear in $\ln \mu$ to all-loop orders, as can be proved by RG consistency
\begin{align}
\label{eq:rg-CS-mu}
\frac{\rd}{\rd\ln\mu }\ln D_1^q(z, b_\perp/z, \xi,\mu) 
=&\, - \gamma^q_{\rm cusp}(\alpha_s(\mu))\ln\frac{\xi}{\mu^2} -\gamma^q_{H} \,.
\end{align}
The $\mu$-evolution  is solved by two conventional Sudakov integrals
\begin{align} \label{eq:Kw_def}
&K_\Gamma (\mu_0, \mu)
=- \intlim{\alpha_s(\mu_0)}{\alpha_s(\mu)}{\alpha_s} \frac{\Gamma(\alpha_s)}{\beta(\alpha_s)}
   \intlim{\alpha_s(\mu_0)}{\alpha_s}{\alpha_s'} \frac{1}{\beta(\alpha_s')}
\,,
\nn\\
&A_\gamma(\mu_0, \mu)
= -\intlim{\alpha_s(\mu_0)}{\alpha_s(\mu)}{\alpha_s} \frac{\gamma(\alpha_s)}{\beta(\alpha_s)}
\,,\end{align}
 their perturbative expansions can be found, e.g.\,, in Ref~\cite{Becher:2006mr}.
 The rapidity
evolution of the TMDs is governed by the Collins-Soper (CS) equation
\begin{align}
\label{eq:rg-CS-nu-1}
K(b_\perp,\mu)\equiv\frac{\rd}{\rd\ln \sqrt{ \xi}}\ln D_1^q (z, b_\perp/z, \xi,\mu)\,,
\end{align}
where $K(b_\perp,\mu)$ is referred to as
the CS kernel.
Again, by RG consistency, the $\mu$-evolution of $K(b_\perp,\mu)$ is controlled by the cusp anomalous dimension 
\begin{align}
\frac{\rd}{\rd\ln\mu} K(b_\perp,\mu) = -2 \gamma^q_{\rm cusp} (\alpha_s(\mu))\,.
\end{align}
From  above we can  solve the RG equation to give
\begin{align}
\label{eq:rg-CS-nu-2}
K(b_\perp,\mu) 
=&-2 A^q_{\rm cusp}(\mu,\mu_b)+\gamma^q_R(\mu_b)\,,
\end{align}
the solution has a evolution factor  $A^q_{\rm cusp}(\mu,\mu_b)$ controlled by the cusp anomalous dimension $\gamma^q_{\rm cusp}$,
as  given  by  Eq.~(\ref{eq:Kw_def}),
it also leaves a residual boundary term $\gamma^q_R(\mu_b)$ known as QCD rapidity anomalous dimension.
While the cusp anomalous dimension $\gamma^q_{\rm cusp}$ is a purely perturbative object which originates from the UV renormalization of the cusped Wilson loops~\cite{Brandt:1981kf,Brandt:1982gz}, the rapidity anomalous dimension $\gamma^q_R(\mu_b)$, on the other hand,  has both  perturbative part as $b_T \to 0$ and  genuine nonperturbative part as $b_T \to \infty$. The perturbative accuracy for the CS kernel has reached  N${}^4$LL~\cite{Li:2016ctv,Moult:2022xzt,Duhr:2022yyp},
and substantial progress has been made toward lattice-QCD determinations for the nonperturbative part as well~\cite{LatticePartonLPC:2022eev,Shanahan:2020zxr,Shanahan:2021tst,Avkhadiev:2024mgd,Schlemmer:2021aij,Li:2021wvl,LatticeParton:2020uhz} .
Eqs.~(\ref{eq:rg-CS-mu}),\,\, (\ref{eq:rg-CS-nu-1}) and (\ref{eq:rg-CS-nu-2})
are referred to as
Collins-Soper equations~\cite{Collins:1981uk,Collins:1987pm,Collins:2011zzd},
they are  equivalent to the modern language of SCET rapidity renormalization groups (RRGs)~\cite{Chiu:2011qc,Chiu:2012ir}.
 Fig.~\ref{fig:EEC-btob} shows the RG-improved prediction for the Mellin space ECs according to the definition  of Eq.~(\ref{eq:mellin-EEC}).
\section{Collinear Limit and the Fragmentation Energy Correlators}
\begin{figure}
\centering
\includegraphics[width=0.25 \textwidth]{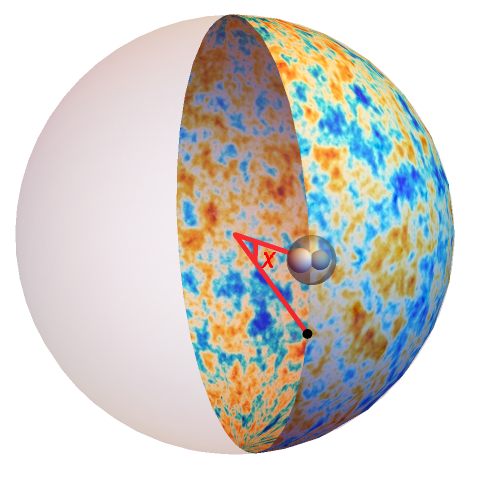}
\caption{ECs collinear limit.}
\label{fig:EEC_cmb_0}
\end{figure}
In this section, we are concerned with collinear limit $\chi\to 0$
where correlation between
the examined hadron and the surrounding radiations is measured,
see Fig.~\ref{fig:EEC_cmb_0}.
This kinematic region is  understood  as the crossed counterpart of target fragmentation region (TFR) in SIDIS,
and the appropriate theoretical framework for describing it is through Fragmentation Energy Correlators (FECs)~\cite{Liu:2024kqt},
whose  operator definitions are given below in a gauge invariant manner
\begin{align}
&\left[D_{h/q}^\text{FEC}\right]^{ij}\left(z,\ln \frac{E_h \sin\chi}{\mu},\mu\right)
\equiv
z
\int \frac{\rd t^-}{2 \pi} e^{i t^-\frac{ P_h^+}{z}}
\nn\\
\times&
\langle 0 | W(\infty\,,t^-)\Psi_i(t^-)
\hat{\mathcal{E}}(\chi)
| h;X\rangle
\langle h;X|
\bar \Psi_j W(0\,,\infty)
  |0\rangle\,,
\nn\\
   \nn\\
  &\left[D_{h/g}^\text{FEC}\right]^{\mu\nu}\left(z,\ln \frac{E_h \sin\chi}{\mu},\mu\right)
  \equiv
  \frac{z^2}{2P_h^+} \int \frac{\rd t^-}{2 \pi} e^{i t^-\frac{ P_h^+}{z}}
    \nn\\
    \times&
  \langle 0 | \mathcal{W}(\infty\,,t^-)
  F^{+\mu}(t^-)
   \hat{\mathcal{E}}(\chi)
  | h;X\rangle
  \langle h;X|
  F^{+\nu}\mathcal{W}(0\,,\infty)
    |0\rangle\,.
\end{align}
The FECs provide a joint description of  parton-to-hadron fragmentation   and 
of its energy correlation pattern with  the  surrounding QCD jets.
In addition, they encode the spin flow in the formation of the observed hadron, analogous  to    
the conventional FFs~\cite{Metz:2016swz}.
In this work, we focus on the unpolarized, chiral-even fragmentation correlators.

To leading power in the collinear limit, 
the EC structure  function in Eq.~(\ref{eq:structure-fun})  is dominated by twist-2 operators with energy flow operators inserted~\cite{Liu:2022wop,Cao:2023qat}, the matching condition reads
\begin{widetext}
 \begin{align}
 \label{eq:fac-x-space}
\frac{\rd  \sigma^{\text{EC}}(x,Q, \chi)}{\rd x} \simeq
&
%
%
 \frac{4\pi \alpha_e^2}{3 Q^2}N_c
\bigg\{\,
\sum_{i=1}^{N_f} Q_i^2 \,
D^{\text{FEC,NS}}_{h/i}(\chi,\mu)
\otimes
\mathscr{C}_{i}^\text{NS}(Q,\mu) + \left(\sum_{i=1}^{N_f} Q_i^2\right)
\nn\\
  \times &
 \bigg[
 \frac{1}{N_f} D^\text{FEC}_{h/q}(\chi,\mu) \otimes \mathscr{C}_{q}^\text{S}(Q,\mu)   
  + D^\text{FEC}_{h/g}(\chi,\mu)\otimes  2\mathscr{C}_{g}^\text{S}(Q,\mu) 
 \bigg]
 \bigg\}
 \,,
 \end{align}
  \end{widetext}
where the nonsinglet and singlet combinations of the FECs are defined as
\begin{align}
D^\text{FEC}_{h/q}(z,\chi)=&\sum_{i=1}^{N_f} D^\text{FEC}_{h/i}(z,\chi)+ D^\text{FEC}_{h/\ib}(z,\chi)\,,
\nn\\
D^\text{FEC,NS}_{h/i}(z,\chi) =&  D^\text{FEC}_{h/i}(z,\chi)+ D^\text{FEC}_{h/\ib}(z,\chi)
\nn\\
-&\frac{1}{N_f}D^\text{FEC}_{h/q}(z,\chi)\,.
\end{align}
The associated energy flow operator $\hat{ \mathcal{E}} (\chi)$ acts only on the collinear sector,
thus, 
 if  $\hat{ \mathcal{E}} (\chi)$ were set to the identity, i.e.\,, in the absence of detectors in the jet fragmentation region (JFR), 
 FECs reduce to standard FFs and the resulting factorization formula coincides with that of  SIA.
 From this perspective, the Wilson  coefficients  are identical to the SIA coefficient functions~\cite{Rijken:1996ns,Mitov:2006wy,He:2025hin},
and the UV renormalization of FECs matches that of the conventional FFs, for instance
\begin{align}
\label{eq:UV-renor}
  \begin{bmatrix}
D^{\text{FEC}}_{h/q}\left(\ln\frac{E_h \sin \chi} {\mu},\mu\right) 
\\
 D^\text{FEC}_{h/g}\left(\ln\frac{E_h \sin \chi} {\mu},\mu\right) 
  \end{bmatrix} 
   =&
     \begin{bmatrix}
D^{\text{FEC}0}_{h/q}\left(\ln\frac{E_h \sin \chi} {\mu},\mu\right) 
\\
 D^{\text{FEC}0}_{h/g}\left(\ln\frac{E_h \sin \chi} {\mu},\mu\right)
  \end{bmatrix}
  \nn\\
   \otimes&\widehat{Z}_\text{S}^\text{FF} 
      \,,
\end{align}
\begin{figure}[htbp]
  \centering
   \subfloat[]{\includegraphics[width=0.24\textwidth]{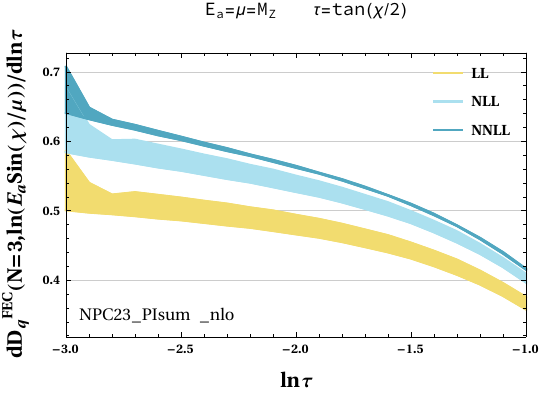}}
   \subfloat[]{\includegraphics[width=0.245\textwidth]{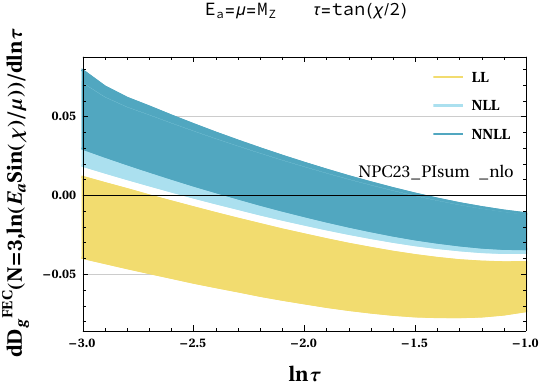}}
 \caption{Mellin moments of FEC singlets renormalized at  fragmenting-parton energy $\mu= E_a=m_Z$. 
 Panel (a) shows the shape of the renormalized quark fragmentation energy correlator, while panel (b) shows the corresponding gluon correlator.}
    \label{fig:SIEC_N3}
\end{figure}
where  the FECs  are and must be parametrized by hadronic logarithm $\ln E_h \sin \chi/\mu$,
$E_h$ being the energy of the examined hadron $h$, in electron-positron annilation $E_h = x\, Q/2$.
The multiplicative renormalization factor $ \widehat{Z}_\text{S}^\text{FF}$  
is the standard FF UV renormalization factor 
for the singlets, the associated 
Dokshitzer-Gribov-Lipatov-Altarelli-Parisi (DGLAP)~\cite{Gribov:1972ri,Lipatov:1974qm,Altarelli:1977zs} evolution kernel reads
 \begin{align}
\widehat{P}_\text{S}(x,\alpha_s) = 
  \begin{pmatrix}
    P_{qq} &  P_{qg}
\\
   2N_f P_{gq} & P_{gg} 
  \end{pmatrix} \,.
\end{align}
Alternatively, one could consider  the  Bjorken-$x$ weighted ECs in Mellin space, defined in Eq.~(\ref{eq:mellin-EEC}).
To this end, we first introduce the Mellin moment of FECs,  
\begin{align}
\mathscr{D}^{\text{FEC}}_{h/i}\left(N,\ln \frac{E_a \sin\chi}{\mu}\right)
=&\int_0^1 \rd z\, z^{N-1} 
\nn\\
\times&
\mathscr{D}^{\text{FEC}}_{h/i}\left(z,\ln \frac{E_a \sin\chi}{\mu}\right)\,.
\end{align}
where the reparametrization  is defined by 
\begin{align}
\mathscr{D}^{\text{FEC}}_{h/i}\left(z,\ln \frac{E_a \sin\chi}{\mu},\mu\right)
=
D^{\text{FEC}}_{h/i}\left(z,\ln\frac{E_h \sin \chi} {\mu},\mu\right) 
\,,
\end{align}
with $E_a=  E_h/z$ as $z$ denotes the momentum fraction of the observed hadron.
In electron-positron annihilation, the energy of the active parton is $E_a= \hat x Q/2 $, where $\hat x= x/z$.
When expressed in terms of the partonic logarithm
$L=\ln(E_a\sin\chi/\mu)$ rather than hadronic variables,
the FECs obey a modified DGLAP equation,
\begin{align}
\label{eq:UV-RG}
&\left.
\frac{\rd}{\rd\ln\mu^2}
\right|_{E_a,\chi}
\mathscr D^{\mathrm{FEC}}_{h/i}(N,L,\mu)
\nonumber\\
&\quad =
\sum_j\int_0^1\rd\xi\,\xi^{N-1}
\mathscr D^{\mathrm{FEC}}_{h/j}
(N,L+\ln\xi,\mu)\,
P_{ji}(\xi,\mu),
\end{align}
where $P_{ji}$ are the ordinary timelike splitting functions.
Unlike ordinary Mellin-space DGLAP evolution, this equation
is nonlocal in $L$: the momentum fraction $\xi$ also shifts
the logarithmic argument of the FEC.

This logarithmic translation is diagonalized by a Fourier
transform,
\begin{equation}
\label{eq:fec-fourier}
\widetilde{\mathscr D}^{\mathrm{FEC}}_{h/i}(N,k,\mu)
=
\int_{-\infty}^{\infty}\rd L\,e^{-ikL}
\mathscr D^{\mathrm{FEC}}_{h/i}(N,L,\mu).
\end{equation}
In Fourier-$k$ space, the shift $L\to L+\ln\xi$ produces
a factor $\xi^{ik}$, changing the Mellin weight from
$\xi^{N-1}$ to $\xi^{N+ik-1}$.
The evolution therefore probes the ordinary timelike
splitting matrix at the shifted complex moment $N+ik$.
Collecting the FECs into a row vector, the solution reads
\begin{equation}
\label{eq:fec-k-solution}
\widetilde{\boldsymbol{\mathscr D}}^{\mathrm{FEC}}_h
(N,k,\mu)
={}
\left(\frac{\mu}{\mu_0}\right)^{ik}
\widetilde{\boldsymbol{\mathscr D}}^{\mathrm{FEC}}_h
(N,k,\mu_0)
\mathbf D^{N+ik}(\mu_0,\mu),
\end{equation}
where
\begin{equation}
\label{eq:fec-evolution-operator}
\mathbf D^M(\mu_0,\mu)
=
\overline{\mathcal T}
\exp\!\left[
\int_{\mu_0}^{\mu}\rd\ln\bar\mu^2\,
\mathbf P(M,\bar\mu)
\right]
\end{equation}
is the ordinary timelike DGLAP evolution operator, with
$\mathbf P(M,\mu)
=\int_0^1\rd\xi\,\xi^{M-1}\mathbf P(\xi,\mu)$.
The ordering symbol $\overline{\mathcal T}$ places later
evolution times to the right, as appropriate for evolution
of a row vector.
The phase in Eq.~\eqref{eq:fec-k-solution} accounts for
$\rd L/\rd\ln\mu^2=-1/2$ at fixed $E_a\sin\chi$.

Defining $L_0=\ln(E_a\sin\chi/\mu_0)$, the inverse Fourier
transform gives the compact representation
\begin{align}
\label{eq:fec-L-operator}
\boldsymbol{\mathscr D}^{\mathrm{FEC}}_h(N,L,\mu)
={}
\boldsymbol{\mathscr D}^{\mathrm{FEC}}_h
\!\left(N,L_0+\overrightarrow{\partial}_N,\mu_0\right)
\mathbf D^N(\mu_0,\mu).
\end{align}
Here the arrow denotes derivatives acting only on the
evolution matrix, not on the explicit $N$ dependence
of the initial FEC; the operator notation is understood
through the inverse Fourier representation.

We implement this evolution numerically at NNLL accuracy
using the NNLO timelike splitting
functions~\cite{Stratmann:1996hn,Almasy:2011eq,Chen:2020uvt}
and the NNLO twist-two matching coefficients derived below.
The resulting numerical solution is shown in
Fig.~\ref{fig:SIEC_N3}. 

\begin{figure}
\centering
\includegraphics[width=0.32 \textwidth]{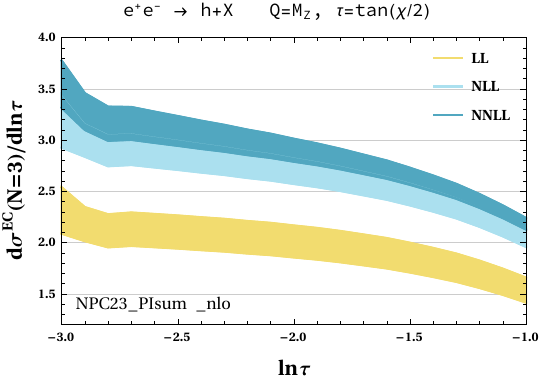}
\caption{NNLL prediction for semi-inclusive energy correlations in the collinear limit.}
\label{fig:EEC-Forward}
\end{figure}

In summary, the Mellin space factorization formula   is
\begin{widetext}
\begin{align}
 \label{eq:fac-N-space}
 \sigma_h^{\text{EC}}(N,Q, \chi)\simeq
&
%
%
\frac{4\pi \alpha_e^2}{3 Q^2}N_c
 \int_0^1 \rd \hat x \, \hat x^{N-1}
\bigg\{\,
\sum_{i=1}^{N_f} Q_i^2 \,
\mathscr{D}^{\text{FEC,NS}}_{h/i}\left(N,\ln\frac{\hat x Q/2 \sin\chi}{\mu},\mu \right)
\mathscr{C}_{i}^\text{NS}(\hat x,Q,\mu) + \left(\sum_{i=1}^{N_f} Q_i^2\right) 
\nn\\
  \times &
   \bigg[
 \frac{1}{N_f} \mathscr{D}^\text{FEC}_{h/q}\left(N,\ln\frac{\hat x Q/2 \sin\chi}{\mu},\mu\right) 
  \mathscr{C}_{q}^\text{S}(\hat x, Q,\mu)   
  + \mathscr{D}^\text{FEC}_{h/g}\left(N,\ln\frac{\hat x Q/2 \sin\chi}{\mu},\mu\right)
    2\mathscr{C}_{g}^\text{S}(\hat x,Q,\mu) 
 \bigg]
 \bigg\}
 \,.
\end{align}
\end{widetext}
The NNLL prediction at a center-of-mass energy corresponding to the $Z$-boson mass scale is shown in Fig.~\ref{fig:EEC-Forward}, 
where we vary the initial boundary and factorization scales around $(\mu_0\,, \mu_f) \sim (Q/2 \sin\chi\,,Q)$.
Unlike the back-to-back jets, where Sudakov suppression arises due to the recoil of soft gluons, 
soft radiation is suppressed by the energy weight in the jet fragmentation region. As a result, the distribution at small angles is not suppressed.
\section{Matching with Collinear Factorization}
The factorization formulas in  Eqs.~(\ref{eq:eec-hadron}), ~(\ref{eq:fac-x-space}) and (\ref{eq:fac-N-space}) do not require a strict scale hierarchy between $q_T \sim Q/2 \sin\chi$ and $\Lambda_\text{QCD}$.
Nevertheless, in the perturbative regime $Q \gg q_T \gg \Lambda_\text{QCD}$, consistency between the TMD/JFR factorization and the standard collinear factorization implies that the TMDs and FECs admit  matching onto conventional FFs.
The perturbative   twist-2 contributions of the TMD fragmentation function have been obtained in~\cite{Luo:2020epw,Ebert:2020qef}.
On the other hand, analogous  to the observation in Refs.~\cite{Liu:2022wop,Cao:2023oef},
 the FECs allow an operator product expansion (OPE) onto energy weighted collinear FFs as follows
\begin{align}
\label{eq:SIEC-OPE}
\mathscr{D}^{\text{FEC}}_{h/i}&\left(z,\ln \frac{E_a \sin\chi}{\mu},\mu\right)
= 
-d_i(z,\mu) +\int_z^1
\frac{\rd \xi}{\xi}
\nn\\
\times\,&
\frac{1}{\xi} d_j(\xi,\mu)
\mathcal{C}_{ji}
\left(
\frac{z}{\xi},
\ln \frac{E_a \sin\chi}{\mu},
\alpha_s(\mu)
\right)
+\mathcal{O}
\left(\frac{\Lambda_\text{QCD}}{q_T}\right)
\,.
\end{align}
The  matching coefficients $\mathcal{C}_{ij}$ is insensitive to the  hadronic state $h$,
thus can be computed from partonic  FECs, and renormalized according to Eq.~(\ref{eq:UV-renor}).
After UV renormalization, the FECs are reparametrized  by the partonic logarithm through 
replacement $\ln \frac{E_h}{\mu}\to \ln \frac{E_a}{\mu} +\ln z$, and matched onto energy densities $1/\xi d_j(\xi,\mu)$.
The renormalization group equation governing the coefficient function follows directly from Eqs.~(\ref{eq:UV-RG}) and (\ref{eq:SIEC-OPE})
\begin{align}
\label{eq:match-rg}
\frac{\rd \mathcal{C}_{ij}}{\rd \ln \mu^2}&
\left(
z\,,
\ln \frac{E_a \sin\chi}{\mu} 
\right)
= 
\int_z^1 
\frac{\rd \xi}{\xi}
\bigg[
\mathcal{C}_{ik}\left(
\frac{z}{\xi}\,,
\ln \frac{\xi E_a \sin\chi}{\mu} 
\right)
\nn\\
\times \,&
 P_{kj}(\xi)
-
 \frac{P_{ik}(\xi)}{\xi}
\mathcal{C}_{kj}\left(
\frac{z}{\xi}\,,
\ln \frac{E_a \sin\chi}{\mu} 
\right)
\bigg]
\,.
\end{align}
The first term in Eq.~(\ref{eq:match-rg}) originates from RG flow with respect to the  UV renormalization scale from above,
while the second term arises from RG flow toward the IR,
the solution  up to NNLO reads
\begin{align}
\label{eq:RG-solution}
\mathcal{C}_{ij}&
\left(
z\,,
\ln \frac{E_a \sin\chi}{\mu} 
\right)
=
\delta_{ij}\delta(1-z)
+
a_s(\mu)
\bigg( 
\mathcal{C}^{(1)}_{ij}(z)
\nn\\
+&
L_q\,
\left(
p_{ij}^{(0)}(z)-P_{ij}^{(0)}(z)
\right)
\bigg)
+
a_s^2(\mu)
\bigg(
\mathcal{C}^{(2)}_{ij}(z)
+
L^2_q
\nn\\
\times
&
\bigg[
\frac{
p_{ik}^{(0)}\otimes p_{kj}^{(0)}(z)
}{2}
+
\frac{P_{ik}^{(0)}\otimes P_{kj}^{(0)}(z)}{2}
-
p_{ik}^{(0)} \otimes P_{kj}^{(0)}(z)
\nn\\
-&
\frac{1}{2}\beta_{0}
\bigg(
p_{ij}^{(0)}(z)-
P_{ij}^{(0)}(z)
\bigg)
\bigg]
+L_q
\bigg[
p_{ik}^{(0)}
\otimes
\mathcal{C}^{(1)}_{kj}
(z)
\nn\\
-
&
\mathcal{C}^{(1)}_{ik}
\otimes
P_{kj}^{(0)}
(z)
-
\beta_0 
\mathcal{C}^{(1)}_{ij}(z)
+
p_{ij}^{(1)}(z)-
P_{ij}^{(1)}(z)
\nn\\
-&
2
\left(p_{ik}^{(0)}-P_{ik}^{(0)}\right)
\otimes
\tilde{P}_{kj}^{(0)}
(z)
\bigg]
\bigg)
+\mathcal{O}(a_s^3(\mu))
\,,
\end{align}
where $a_s(\mu)=\alpha_s(\mu)/(4\pi)$ and $L_q=2 \ln \frac{E_a \sin\chi}{\mu}$.
The modified splitting functions are given by
\begin{align}
p_{ij}^{(m)}(z)=1/z P_{ij}^{(m)}(z)\,,\quad 
\tilde{P}_{ij}^{(0)}(z)=\ln z \,P_{ij}^{(0)}(z)\,.
\end{align} 
The results above serve as boundary conditions for the NNLL resummation for the semi-inclusive energy correlations.
They also directly yield the collinear limit of inclusive EEC,
as the latter  is given by
\begin{align}
\sum_{\{A\neq B\}} \frac{E_A }{Q/2} \frac{E_B }{Q/2}=
\frac{1}{2}  \sum_{A} \left(\frac{E_A }{Q/2}\right)^2 \sum_{B \neq A} \frac{E_B}{E_A}\,.
\end{align}
As a result, taking the third Mellin moment of Eq.~(\ref{eq:fac-N-space}) and summing over all hadron species
including the EEC contact terms
 yields the collinear factorization formula for  inclusive EEC,
 \begin{widetext}
\begin{align}
 \label{eq:fac-EEC-colli}
 \sigma^{\text{EEC}}(Q\,,\chi)\simeq
\sigma_0^\text{tot}
\sum_{i=q,g}
 {\mathcal J}_{i}\left(\ln \frac{Q/2 \sin\chi}{\mu}+\cdot\,,\mu\right) 
 \odot
  \mathscr{C}_{i}^\text{S}(\cdot\,, Q\,,\mu)  
  +\mathcal{O}
\left(\frac{\Lambda_\text{QCD}}{q_T}\right)
 \,.
\end{align}
\end{widetext}
 with the contact contributions canceling against the EEC contact terms, as demonstrated below.
Here, $\sigma_0^\text{tot}=\sum_{f=1}^{N_f} \sigma^{(0)}_{f \bar f}\,,\,\, 
 \sigma^{(0)}_{f \bar f}=  \frac{4\pi \alpha_e^2}{3 Q^2}N_c Q_f^2$.
 Note that the FEC nonsinglets do not contribute to inclusive EEC in the collinear limit.
 Furthermore, the  convolution is defined by 
 \begin{align}
& {\mathcal J}_{i}\left(\ln \frac{Q/2 \sin\chi}{\mu}+\cdot\,,\mu\right) 
 \odot
  \mathscr{C}_{i}^\text{S}(\cdot\,, Q\,,\mu)  =
  \nn\\
 & \int_0^1 \rd \hat x \,{\hat x}^{2} \,
   {\mathcal J}_{i}\left(\ln \frac{\hat x Q/2 \sin\chi}{\mu}\,,\mu\right) 
     \mathscr{C}_{i}^\text{S}(\hat x\,, Q\,,\mu)\,.
 \end{align}
 Owing to the sum rule of collinear FFs, 
 the collinear EEC jet function  is given by the sum of third Mellin moment of  FEC twist-2 matching coefficients, 
 reproducing the result of Refs.~\cite{Dixon:2019uzg,Dixon:2018qgp}
 \begin{align}
  {\mathcal J}_{i}\left(\ln \frac{E_a \sin\chi}{\mu}\,,\mu\right) 
  =\sum_j 
   \mathcal{C}^\text{S}_{ji}\left(N=3\,,\ln \frac{E_a \sin\chi}{\mu}\,,\mu\right)\,.
 \end{align}
 Our analysis is consistent with dihadron fragmentation framework~\cite{Lee:2025okn,Kang:2025zto,Chang:2025kgq}
 where  the contact term of the dihadron fragmentation function contributes to the EEC collinear limit by
  \begin{align}
  \label{eq:dihadron}
 &\frac{1}{2}\sigma_0^\text{tot}
 \sum_{h'}
 \int_0^{1-x}
 \frac{x'  \rd x'}{x}
 \int_{x+x'}^1
\frac{\rd z}{z^2}
 D_{i\to h h'}\left(\frac{x}{z},\frac{x'}{z}\right)
 \mathscr{C}_{i}(z,Q)
 \nn\\
 =&
 \frac{1}{2}\sigma_0^\text{tot}
\int_x^1\frac{\rd z}{z}  \left(\frac{1}{z}-1\right) d_i(z,\mu)  \mathscr{C}_{i}\left(\frac{x}{z},Q,\mu\right)\,,
 \end{align}
 where $x'$ denotes the momentum fraction of  another hadron~\cite{Konishi:1979cb,Vendramin:1981te,Pitonyak:2025lin,Rogers:2024nhb,Pitonyak:2023gjx},
 and for convenience we  slightly adjust the normalization of the SIA coefficient functions in Eq.~(\ref{eq:fac-x-space}) so that the convention matches those used in Refs.~\cite{Rijken:1996ns,Lee:2025okn}. 
To the  EEC observable, 
the first term $d_i(z,\mu) /z$  in Eq.~(\ref{eq:dihadron})  gives an  IR finite contribution 
while the second term matches the first term of the collinear expansion in Eq.~(\ref{eq:SIEC-OPE}), and
yields an IR-divergent contribution 
 \begin{align}
 -\frac{1}{2}\sigma_0^\text{tot}\sum_h \sum_i d_i^h(N=3,\mu)\mathscr{C}_{i}^\text{S}(N=3,Q\,,\mu)\,,
 \end{align}
 which exactly cancels with the contact term contribution in the EEC, 
 thereby ensuring the IR safety of the EEC observable. 
 The factor of $1/2$ accounts for the normalization \(\sum_{h\in X}z_h=2\) in each final-state event, as required by energy conservation,
 as in Eq.~(2.18) of Ref.~\cite{Rijken:1996ns} for the inclusive cross section.
 After isolating the contact term, the remaining contributions from the dihadron fragmentation functions arise entirely from the hard-collinear region~\cite{Cao:2023oef}.
\begin{figure}
\centering
\includegraphics[width=0.33 \textwidth]{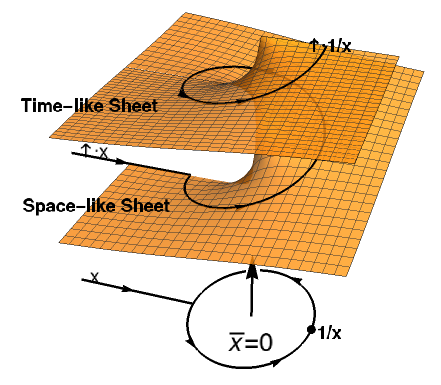}
\caption{The inversion relation between Bjorken variable $x_B$ and Feynman variable $z_F=1/x_B$.}
\label{fig:NEEC_cross}
\end{figure}

As a concluding remark in this section, 
we point out that the holomorphic parts of  bare partonic FECs are obtained from crossings of Nucleon Energy Correlators (NECs)~\cite{Liu:2022wop,Cao:2023qat}.
The crossing relations differ from their TMD counterparts by an overall minus sign due to the energy weight~\cite{Chen:2020uvt}.
For the quark FECs/NECs, the crossing relation is 
\begin{gather}
   D^\text{FEC}_{\bar \imath/\bar{q}}(x, E_P, \chi) = (-1)^{1+i_F}
 x^{1-2 \e}
F^\text{NEC}_{q/i} \left( - \frac{e^{i\pi}}{x}, -E_P, \chi \right) \,,
\label{eq:rulea}
\end{gather}
where $i_F = 1$ if $i$ is a fermion, and $0$ if boson. The minus sign originates from crossing a fermion from initial state to final state.
For the gluon FECs/NECs, the crossing relation is
\begin{align}
   D^\text{FEC}_{\bar \imath/g}(x, E_P, \chi) =  (-1)^{i_F}
 x^{1-2 \e}
F^\text{NEC}_{g/i} \left( - \frac{e^{i\pi}}{x}, -E_P, \chi \right) \,.
\label{eq:ruleb}
\end{align}

The exact meaning of the inversion $x\to1/x$  is illustrated in Fig.~\ref{fig:NEEC_cross}.
In the first step, a representative path from $x\to 1/x$ is drawn  in the base space:
it first approaches the neighborhood of threshold branch point $\bar x=1-x=0$,
encloses it counterclockwise via the map $I=[0,1] \overset{\omega}{\mapsto} \mathbb{S}^1$ where $\omega(x)=e^{2\pi i x}$
 and then proceeds to destination $1/x$ without enclosing any further branch points.
In the second step, the path is  lifted to its covering space, taking the spacelike NECs
$F^\text{NEC}_{j/i} ( x, \dots )$ as the initial value of the lift,  which is unambiguously given since it is defined in the  Euclidean region.
Owing to the unique lifting property, the timelike FECs are obtained  by directly computing the 
monodromy of individual  functions. The complete analytic expressions for the coefficient functions are provided in the ancillary material~\cite{ancfiles}, along with numerical routines implemented in \texttt{PolyLogTools}~\cite{Duhr:2019tlz}.
The crossing relation between dynamics of  TFR and JFR even persists beyond the paradigm of amplitudes,
in  Refs~\cite{Basso:2006nk,Caron-Huot:2022eqs,Chen:2020uvt}, a reciprocity relation   was proposed  between  spacelike  and timelike anomalous dimensions of twist-2 operators. However, a  rigorous proof in QCD is still lacking and remains an open problem  for future investigations. 
\section{Summary and Outlook}
In this work, we investigate energy correlations for single-inclusive hadron production in electron-positron annihilation via photon decay. Within SCET, we demonstrate factorization into universal nonperturbative correlation functions  in both  small- and wide-angle limits. We compute the twist-2 matching of FECs at NNLO and achieve joint $\mathrm{N}^3$LL/NNLL renormalization-group-improved predictions in the Sudakov and jet fragmentation regions; results consistent with previous findings~\cite{Moult:2018jzp,Dixon:2018qgp,Dixon:2019uzg}.
These high-precision predictions provide the theoretical ingredients for precision extractions of TMD fragmentation functions in the Sudakov region and improve theoretical control over parton-to-hadron transition  in the jet fragmentation region.

Looking ahead, several further improvements are worth pursuing.
First, extending our framework to the $Z$ pole while retaining the angular dependence relative to the lepton beam would enable precision studies of parity-violating forward-backward asymmetries. In particular, heavy-flavor charge-tagged energy correlators could provide new insights into QCD corrections to the bottom-quark forward-backward asymmetry, whose long-standing tension with the Standard Model motivates precision measurements at future high-luminosity $e^+e^-$ colliders, such as CEPC~\cite{CEPCStudyGroup:2018ghi} and FCC-ee~\cite{FCC:2018evy}.
Second, the fixed-order Mellin moments develop singularities at \(N=2\) due to enhanced small-\(z\) logarithms, motivating an all-order resummation in this regime.
Third, extending the current framework to chiral-odd operators, thereby enabling probes of quark transversity~\cite{Cao:2025icu}, represents an interesting direction for ongoing work.
\section*{Acknowledgments}
I thank Tong-Zhi Yang and Xiaohui Liu for useful discussions.
\appendix 
\section{Small-x logarithms}
\label{sec:small-x}
In this appendix, we collect the small-$x$ logarithms of the twist-2 matching coefficients for the Fragmentation Energy Correlators.
The complete analytic expressions of the coefficient functions are provided in the ancillary material.
\begin{align}
z^2\mathcal{C}^{(1)}_{qq}(z)\simeq& 0\,, \quad 
z^2\mathcal{C}^{(1)}_{gq}(z)\simeq 8 C_F \ln z\,,
\nn\\
z^2\mathcal{C}^{(1)}_{qg}(z)\simeq& 0\,, \quad 
z^2\mathcal{C}^{(1)}_{gg}(z)\simeq 8 C_A \ln z\,,
\end{align}
\begin{widetext}
\begin{align}
z^2\mathcal{C}^{(2)}_{qq}(z)\simeq 
\frac{32}{3} C_F N_f \ln^2 z
+
\frac{16}{3} C_F N_f \ln z
+C_F N_f \left(
-\frac{296}{27}
+\frac{16}{3}\zeta_2
\right)\,,
\end{align}
\begin{align}
z^2\mathcal{C}^{(2)}_{gq}(z)\simeq &
-\frac{80}{3}C_A C_F\ln^3z-\frac{212}{3}C_A C_F\ln^2z
+
\left(
-\frac{736}{9}C_F^2
+C_A C_F
\left(\frac{1204}{9}-32\zeta_2\right)
\right)
\ln z
\nn\\
+&C_A C_F\left(
\frac{4424}{27}
-\frac{160}{3}\zeta_2
-120 \zeta_3
\right)\,,
\end{align}
\begin{align}
z^2\mathcal{C}^{(2)}_{qg}(z)\simeq 
\frac{32}{3} C_A N_f \ln^2 z
+
\frac{16}{3} C_A N_f \ln z
+C_A N_f \left(
-\frac{296}{27}
+\frac{16}{3}\zeta_2
\right)\,,
\end{align}
\begin{align}
z^2\mathcal{C}^{(2)}_{gg}(z)\simeq&
-\frac{80}{3}C_A^2\ln^3z
+\left(
-\frac{220}{3}C_A^2-\frac{8}{3}C_AN_f+\frac{16}{3}C_F N_f
\right)\ln^2z
+\left(
-\frac{92}{9}C_A N_f
+C_A^2\left(
\frac{536}{9}-32\zeta_2
\right)
\right)\ln z
\nn\\
+ &C_A N_f\left(
\frac{562}{27}-\frac{16}{3}\zeta_2
\right)
+
C_F N_f
\left(
-\frac{1124}{27}
+\frac{32}{3}\zeta_2
\right)
+C_A^2
\left(
\frac{1616}{9}
-\frac{176}{3}\zeta_2
-120\zeta_3
\right)\,.
\end{align}
\end{widetext}

\bibliographystyle{apsrev4-1} 
\bibliography{SIA_EEC} 

\allowdisplaybreaks

\end{document}